\documentclass[twocolumn]{aastex63}

\usepackage{amsmath}
\usepackage{amssymb}
\usepackage{euscript}
\usepackage{bbold}
\usepackage{paralist}
\usepackage{color}
\usepackage{amsmath}
\usepackage{lineno}
\usepackage{lineno}
\usepackage{breqn}

\usepackage{comment}

\shorttitle{False Alarm Rate}
\shortauthors{Soumagnac et al.}

\begin{document}


\title{Fast and accurate false-alarm probabilities for linear detection statistics in Poisson noise}


\correspondingauthor{Maayane T. Soumagnac}
\email{mtsoumagnac@lbl.gov}
\author[0000-0001-6753-1488]{Maayane T. Soumagnac}
\affiliation{Department of Physics, Bar-Ilan University, Ramat-Gan 52900, Israel}
\affiliation{Lawrence Berkeley National Laboratory, 1 Cyclotron Road, Berkeley, CA 94720, USA}
\author[0000-0002-6786-8774]{Eran O. Ofek}
\affiliation{Benoziyo Center for Astrophysics, Weizmann Institute of Science, 76100 Rehovot, Israel}
\author{Shachar Israeli}
\affiliation{Department of Physics, Bar-Ilan University, Ramat-Gan 52900, Israel}
\author{David Glukhov}
\affiliation{Department of Physics, Bar-Ilan University, Ramat-Gan 52900, Israel}


\begin{abstract}

Detecting weak signals in noisy data often requires calculating the statistical significance, or false-alarm probability, of a detection statistic. When the underlying noise follows Poisson statistics, as in low-count measurements such as photon-starved astronomical images, this calculation becomes non-trivial, and existing approaches typically rely on extensive simulations to estimate detection thresholds. We derive an accurate and efficient solution for computing false-alarm probabilities for arbitrary linear detection statistics under Poisson noise. We show that the detection statistic is a {\it compound Poisson random variable}, 
which unlocks efficient implementations based on the Fast Fourier Transform (FFT) and the Panjer recursion. The method is exact up to numerical discretization, accelerates computation by several orders of magnitude compared with simulations, and enables evaluation of false-alarm probabilities 
computationally inaccessible through simulation. Although motivated by low-count astronomical imaging, the framework applies broadly to detection problems involving Poisson-distributed events. It transforms the computation of detection significance under Poisson noise from a simulation problem into a deterministic numerical calculation. We provide open-source Python and MATLAB implementations.

\end{abstract}


\section{Introduction}

In astronomy and in other fields of science, the detection of a signal (e.g. an astronomical source in an image), can be formulated as a hypothesis testing problem. The goal is to discriminate between the null hypothesis $\mathcal{H}_0$ (corresponding to the absence of a source) and the alternative hypothesis $\mathcal{H}_1$ (corresponding to the presence of a source). In this context, the significance of the detection can be expressed in terms of the false alarm rate $\gamma$: how often are we willing to reject $\mathcal H_0$ (i.e. to declare a detection) while $\mathcal H_0$ is actually true (i.e. there is no source).

In order to declare a detection, we need a detection statistic $S$  (which is calculated from the data) and a threshold $S_{\rm th}$: if $S \ge S_{\rm th}$ we declare a detection. The threshold $S_{th}$ is related to the false-alarme rate $\gamma$ through:
\begin{equation}\label{eq:gamma}
\gamma = P(S \ge S_{\rm th}) = \int_{S_{\rm th}}^{\infty} P(S|\mathcal{H}_0)\,dS\,.
\end{equation}

Often, the statistic $S$ is the result of a linear filtering operation, a weighted sum (or linear combination) of nearby data values:
\begin{equation}\label{eq:filter}
S(q_0)=\sum_q \mathcal{K}(q-q_0)M(q),
\end{equation}
where $M$ denotes the data/measurements (of any dimension) values, $q$ denotes the (multi-dimensional) coordinates in $M$, $q_{0}$ is a specific coordinate in which we calculate the statistics and $\mathcal{K}$ is called the {\it filter}, or the {\it kernel}. 
This equation is simply the definition of cross-correlation (i.e., a convolution with coordinate reversal). A nice feature of it is that, in Fourier space, convolution becomes multiplication and can be calculated efficiently using the Fast Fourier Transform (FFT).

A particularly common application of linear filtering is {\it matched filtering}, in which the shape of the signal we wish to detect is known.
In the case of Gaussian, spatially uniform, independent noise, the Neyman--Pearson lemma \citep{Neyman1933} dictates the choice of the kernel $\mathcal{K}$: the optimal statistic is the cross-correlation of the data with the expected signal. For example, when searching for a point source in an astronomical image, the optimal detection statistic is obtained by filtering the image with the Point Spread Function (PSF; \citealt{Stetson1987_DAOPHOT,Schechter+1993_DoPHOT,Ofek+2023PASP_LAST_PipeplineI}). Conveniently, in the Gaussian case, the corresponding threshold $S_{\rm th}$ can also be calculated analytically
(e.g., \citealt{Zackay+2017_CoadditionI,Zackay+2016_ZOGY_ImageSubtraction}).

In the low-count regime, however, where the background is intrinsically Poissonian, the Gaussian approximation is no longer valid and the problem becomes more complicated. Two challenges arise in this regime: (1) the choice of a kernel $\mathcal{K}$ that yields an optimal detection statistic $S$, and (2) the calculation of the threshold $S_{\rm th}$ above which a detection can be declared. In astronomy, this regime is particularly relevant for source detection in low-count X-ray images, as well as in UV, $\gamma$-ray, and neutrino detectors.

Regarding the choice of the kernel $\mathcal{K}$, and in the context of X-ray astronomy, different astronomical source-detection pipelines have adopted different detection kernels. For example, in XMM-Newton, the {\tt eboxdetect} algorithm uses a sliding-box (top-hat) filter to produce an initial list of source candidates \citep{Loaring2005}, such that $\mathcal{K}$ is unity inside the detection box and zero outside.
This is followed by an accurate maximum likelihood fit.
The same principle is used in the more recent fourth XMM-Newton serendipitous source catalog, with a $20''\times20''$ detection box \citep{Webb2020}. Similarly, in eROSITA, source candidates are identified using the {\tt erbox} sliding-box algorithm applied to smoothed images \citep{Brunner2022}. In contrast, the Chandra {\tt wavdetect} algorithm \citep{Freeman2002} uses Mexican-hat wavelets as detection kernels \citep{Freeman2002}.

Beyond these practical choices, the question of which kernel is optimal for source detection under Poisson noise has also been investigated theoretically. \citet{Stewart2006} estimated an optimal convolution kernel, while \citet{Ofek+2018_PoissonNoiseMF} derived the Neyman--Pearson optimal detection statistic analytically and showed that it can itself be expressed as a linear filter, with kernel
\begin{equation}\label{eq:kpmf}
\mathcal{K}
=
\ln\left(1+\frac{F}{B}P\right),
\end{equation}
where $P$ is the PSF, $F$ is the source flux assumed under the alternative hypothesis, and $B$ is the background level (e.g., counts per pixel).

Once the kernel $\mathcal{K}$ is chosen, the second challenge is to determine what constitutes a statistically significant detection, i.e., to calculate the detection threshold $S_{\rm th}$ corresponding to a desired false-alarm probability. Simple top-hat/sliding-box filters, such as those used by {\tt eboxdetect} in XMM-Newton and {\tt erbox} in eROSITA, have an advantage: the sum of independent Poisson counts within the detection box is itself Poisson, so its detection significance can be calculated analytically. However, such filters are sub-optimal and therefore sacrifice detection sensitivity. This limitation is mitigated in these pipelines by adopting a permissive threshold at the initial candidate-detection stage and subsequently applying a more stringent criterion based on maximum-likelihood PSF fitting (which is equivalent for using the matched filtering with Equation~\ref{eq:kpmf}).
Conversely, more informative weighted filters, such as the wavelet used by {\tt wavdetect} or the optimal filter of \citet{Ofek+2018_PoissonNoiseMF}, exploit spatial information about the expected signal, but their output no longer follows a simple Poisson distribution, making the corresponding false-alarm probability substantially more difficult to calculate.

In the absence of an analytical expression for the null distribution of $S$, a natural approach is to generate a large number of simulated background-only images. Each simulated image is filtered with the kernel, generating samples of the statistic $S$ under the null hypothesis, from which $P(S\mid\mathcal{H}_0)$ can be estimated. The threshold $S_{\rm th}$ can then be determined from Equation~\ref{eq:gamma}. This is what was done to determine the significance of the statistic used by {\tt wavdetect}, using $\sim 50,000$ $1024\times1024$ Poisson background images. Extensive simulations are also used in XMM-Newton and eROSITA to calibrate the performance of their source-detection pipelines \citep{Loaring2005,Liu2022,Brunner2022}.

The main limitation of this simulation-based approach is its computational cost, which increases (1) linearly with the size of the filter and (2) inversely with the target false-alarm probability $\gamma$. The latter limitation is particularly severe: at least $\mathcal{O}(\gamma^{-1})$ simulations are required to obtain even one expected realization in the relevant tail\footnote{For $N_{\rm sim}$ independent simulations under $\mathcal{H}_0$, the expected number of realizations exceeding a threshold with false-alarm probability $\gamma$ is $N_{\rm sim}\gamma$, so that $N_{\rm sim}\sim\gamma^{-1}$ is required for this expectation to reach unity.}.

For example, in the case of {\tt wavdetect}, the number of simulations performed allowed the null distribution to be calibrated down to a false-alarm probability of approximately $10^{-7}$, corresponding to about $5.2\sigma$ for a one-sided Gaussian probability.

Here, we derive an analytical approach for calculating false-alarm probabilities in the presence of Poisson noise that is independent of the choice of linear filter. The solution we present enables the detection threshold $S_{\rm th}$ to be calculated several orders of magnitude faster than with simulation-based approaches, and at significance levels and filter sizes that would otherwise be computationally inaccessible. 

 In \S\ref{sec:compound_solution_section}, we show that the detection statistic meets the definition of a {\it compound Poisson variable}, a type of random variable whose probability distribution is known and has been extensively studied in a very different context: the field of risk theory (for the calculation of the  aggregate loss distributions). Once this is established, we can make use of existing methods to compute this probability distribution, two of which are presented in \S\ref{sec:method}. We also present, in the Appendix \ref{appendix:enumeration}, a different approach based on the enumeration of events, which is equivalent to the compound variable formalism derived in \S\ref{sec:compound_solution_section}, but is computationally heavier. We compare the different approaches and implementations in \S\ref{sec:tests}, provide the corresponding code in \S\ref{sec:code}, and present our conclusions in \S\ref{sec:conclusion}.


We note that the calculation we propose applies to the per-resolution element false-alarm rate, and we do~not account for the look-elsewhere effect, which should be quantified in order to determine the global false-alarm rate (e.g., \citealt{Vio2018_MatchedFilter_LooksElseWhere}).

\section{The detection statistic $S$ is a compound Poisson random variable}
\label{sec:compound_solution_section}

%
\subsection{Definition of a compound Poisson random variable}\label{sec:compound_def}
In probability theory, a {\it compound Poisson random variable} $S$ is the sum of independent identically-distributed (i.i.d.) random variables, where the number of terms to be added is itself a Poisson-distributed variable:

\begin{equation}\label{eq:compound_def}
S=\sum_{i=1}^{N}X_i,\qquad N\sim {\rm Poisson}(\Lambda).
\end{equation}
Here, $\sim$ denotes ``distributed as'', and $N$ is independent of the $X_i$.

In the rest of this section, we show that under $\mathcal{H}_0$, the detection statistic $S$ defined
by Equation~\ref{eq:filter} meets the definition of a compound Poisson variable. The advantage of this formulation
is that the probability distribution of a compound Poisson variable is well
known: it can be expressed as a Poisson-weighted sum of convolution powers of
a single, easily computed distribution
\citep{Feller1968,Schilling2016}.

In what follows, for convenience, we use the terminology of astronomical
images: an ``event'' is a photon hitting a pixel within the kernel support,
the data $M$ are an image in units of counts (and in the case of a background-only image, ), and $M_{\rm b}(q)$ is the number of
photons hitting pixel $q$ in a background-only image. The derivation, however, applies more generally
to any setting in which independent Poisson-distributed events contribute
to a linear statistic.


\subsection{Two possible representations}\label{sec:two_approaches}

We aim to calculate the probability distribution of the statistic $S$ under the null hypothesis $\mathcal{H}_0$, which we denote by
\begin{equation}
\mathcal{P}_{\mathcal{H}_0}(s)\equiv P(S=s\mid \mathcal{H}_0).
\end{equation}
Under $\mathcal{H}_0$, there is no source at the position $q_0$, and the image is described by background noise only: we denote it as $M_{\rm b}(q)$. The background photons are modeled as a Poisson point process: we assume that the counts in different pixels are independent Poisson random variables,
\begin{equation}\label{eq:back}
M_{\rm b}(q)\sim {\rm Poisson}\left[\lambda(q)\right].
\end{equation}
In the case of a spatially uniform background\footnote{At least locally, over a region larger than the support of the kernel}, the expected background count is the same in every pixel, so that $\lambda(q)=B$,
where $B$ is the mean background count per pixel.

Since the values of $S(q_0)$ at different positions $q_0$ are generally correlated, we derive the distribution for one fixed position. Without loss of generality, we choose $q_0=0$, so that
\begin{equation}\label{eq:filter_simple}
S=\sum_{q=1}^{N_{\rm pix}} M_{\rm b}(q)\mathcal{K}(q).
\end{equation}

where $M_{\rm b}(q)$ is the number of counts measured in pixel $q$ in the background-only image.

A realization in which $n$ photons hit $N_{\rm pix}$ pixels can be
described in two equivalent ways:


\begin{enumerate}
\item One possible approach is to view $M_{\rm b}$ as a collection of pixels with
various levels of occupancy. The occupancy vector
$\boldsymbol{M_{\rm b}}=(M_{\rm b,1},M_{\rm b,2},\ldots,M_{\rm b,N_{\rm bpix}})$, where
$M_{\rm b,j}\equiv M_{\rm b}(q_j)$ is the random number of counts in pixel $j$ located at
position $q_j$, records the number of photons in \textit{each pixel}.
In order to calculate $\mathcal{P}_{\mathcal{H}_0}$, we can consider all
possible realizations of the occupancy vector and calculate the probability
of each realization using combinatorics. We develop this solution in
Appendix~\ref{appendix:enumeration}.

\item An equivalent approach, adopted in this section, is to view $M_{\rm b}$ as a
collection of the individual events that generate it. The vector of photon
positions
$\boldsymbol{Q}=(Q_1,\ldots,Q_n)$ records the pixel position hit by
\textit{each photon}. Here, $Q_i$ is a random variable that can take any of
the pixel positions $q_1,\ldots,q_{N_{\rm pix}}$; for example, $Q_i=q_j$
means that photon $i$ falls in pixel $j$. Several vectors of photon positions
$\boldsymbol{Q}$ may correspond to the same realization of the occupancy
vector $\boldsymbol{M_{\rm b}}$, differing only in the ordering of the photons.
\end{enumerate}

\begin{figure*}
    \centering
    \includegraphics[width=0.9\linewidth]{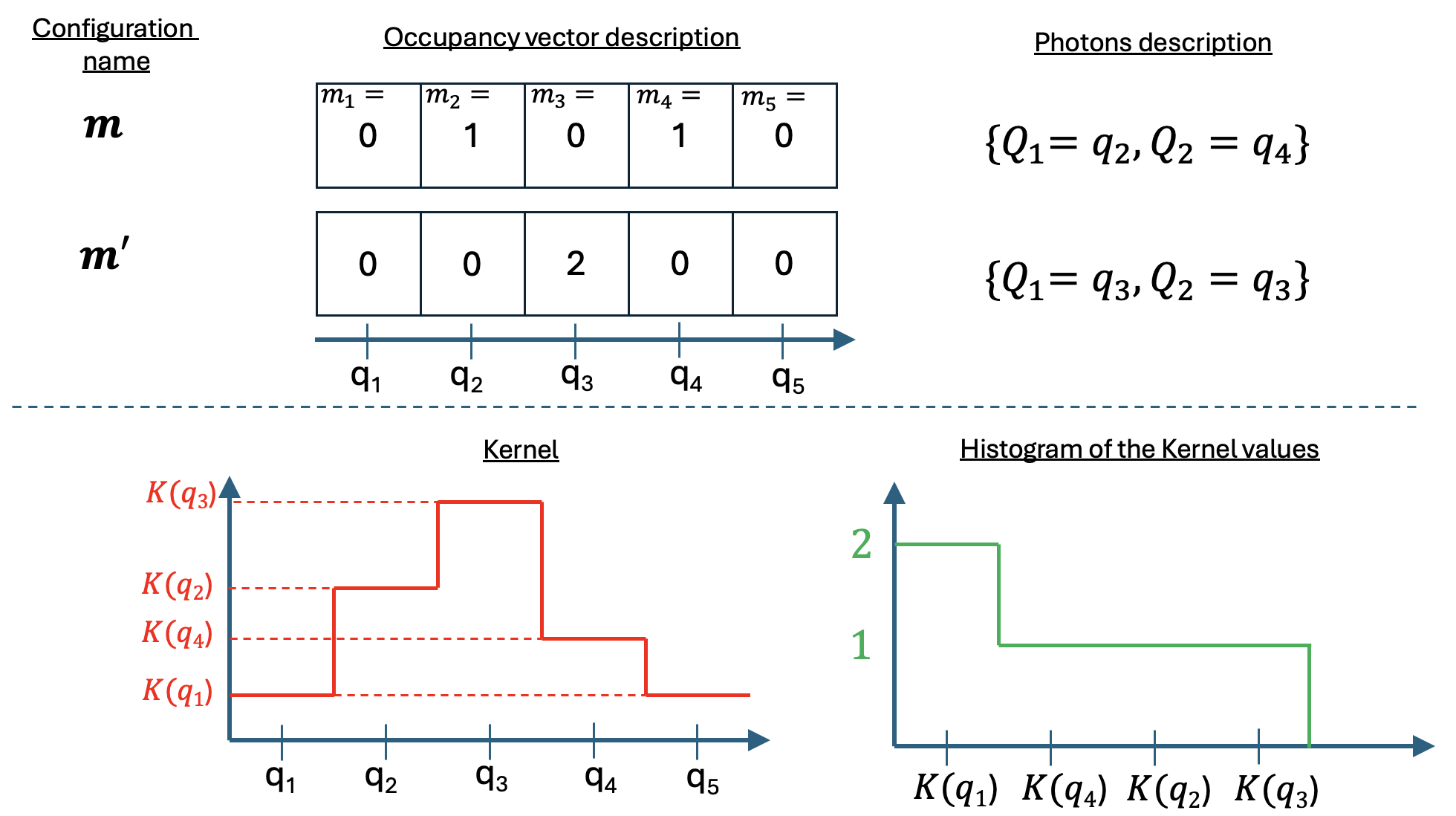}
    \caption{{\bf A Simple Example}. In the top table, we show two realizations $\boldsymbol{m}$ and $\boldsymbol{m'}$ of the occupancy vectors $\boldsymbol{M}$ and $\boldsymbol{M'}$, corresponding to $N_{\rm pix}=5$ and $N=2$. The middle column shows the occupancy-vector representation of each realization: in the realization $\boldsymbol{m}$ we have $m_2=1$ and $m_4=1$ and zeros elsewhere., whereas in the realization $\boldsymbol{m'}$ we have $m_3=2$ and $0$'s everywhere else. An alternative description is shown in the right column, where instead of recording the number of photons in {\it each pixel}, we record the position $Q_i$ hit by {\it each photon}. In the realization $\boldsymbol{m}$, the first photon hit the position $Q_1=q_2$ and the second photon hit the position $Q_2=q_4$, whereas in the realization $\boldsymbol{m'}$, the first photon hit the position $Q_1=q_3$ and the second photon also hit the position $Q_2=q_3$. The lower left plot shows an example of a kernel, and the lower right plot shows the histogram of the values of this kernel ($\mathcal{K}(q_1)=\mathcal{K}(q_5)$ and therefore it appears twice in the histogram, while $\mathcal{K}(q_2)$, $\mathcal{K}(q_3)$ and $\mathcal{K}(q_4)$ each appear once). This histogram can also be written as $\sum_{j=1}^{N_{\rm pix}}\delta\!\left(x-\mathcal K(q_j)\right)$ and yields $p_X(x)$, the probability distribution of the random variables $X_i$, when multiplied by $\frac{1}{N_{\rm pix}}$.}
    \label{fig:illustration}
\end{figure*}

Figure~\ref{fig:illustration} illustrates the equivalence between these two
descriptions. In the remainder of this section, we adopt the second approach. In Appendix~\ref{appendix:equivalence} we demonstrate formally the equivalence between these two representations.

\subsection{$S$ is a compound Poisson random variable}\label{sec:compound_proof}

\subsubsection{$S$ is a sum of random variables}
Let $N$ be the total number of background
photons within the support of the kernel and let $Q_i$ denote the
position hit by photon $i$. For any realization in which $N=n$, the
number of events in pixel $q$ can be written exactly as
\begin{equation}
M_{\rm b}(q)
=
\sum_{i=1}^{n}
\mathbb{1}_{\{Q_i=q\}},
\label{eq:M_events}
\end{equation}
where $\mathbb{1}_{\{Q_i=q\}}$ is the indicator function, equal to
unity when its argument is true (i.e. here, when photon $i$ falls in pixel $q$) and zero otherwise.

Substituting Equation~\ref{eq:M_events} into
Equation~\ref{eq:filter_simple} gives
\begin{align}
S
&=
\sum_q
\mathcal{K}(q)
\sum_{i=1}^{n}
\mathbb{1}_{\{Q_i=q\}}
\nonumber\\
&=
\sum_{i=1}^{n}
\sum_q
\mathcal{K}(q)
\mathbb{1}_{\{Q_i=q\}}
\nonumber\\
&=
\sum_{i=1}^{n}
\mathcal{K}(Q_i).
\label{eq:S_events}
\end{align}
Thus, although Equation~\ref{eq:filter_simple} is expressed as a weighted
sum of {\it pixel} counts, it can equivalently be viewed as a sum of the
individual contributions of the {\it photons} falling within the support of the kernel. 

If we define the contribution to $S$ of a single photon as the random
variable
\begin{equation}
X_i
\equiv
\mathcal{K}(Q_i), 
\label{eq:X_def}
\end{equation}
we can write $S$ as the sum of random contributions from each photon:

\begin{equation}\label{eq:sum}
S=\sum_{i=1}^{n} X_i\,.
\end{equation}

\subsubsection{The number of terms in the sum is Poisson-distributed}
We now show that the number of terms in the sum of Equation~\ref{eq:sum} is itself Poisson distributed (another ingredient in the definition of a compound Poisson random variable).
In our physical model, background photons are modeled as a Poisson point process over the filter support. The expected number of photons in pixel $q$ is $\lambda(q)$ (see also Equation~\ref{eq:back}). The total number of background photons within the kernel support is
\begin{equation}
N
=
\sum_{q=1}^{N_{\rm pix}} M_{\rm b}(q)\;,
\end{equation}
and since the counts $M_{\rm b}(q)$ in the different pixels are independent Poisson
random variables with means $\lambda(q)$, their sum is also Poisson distributed,
\begin{equation}\label{eq:poiss}
N
\sim
{\rm Poisson}(\Lambda),
\qquad
\Lambda=\sum_{q=1}^{N_{\rm pix}} \lambda(q)\,
\end{equation}
(with $\Lambda=N_{\rm pix}B$, in the uniform background case). Thus, the probability of observing $n$ photons within the kernel support is
\begin{equation}
P(N=n)
\equiv
P_{\rm pois}(n)
=
\frac{\Lambda^n e^{-\Lambda}}{n!}.
\end{equation}



\subsubsection{The $X_i$'s are i.i.d}

Conditional on $N=n$, the photon positions \(Q_1,\ldots,Q_n\) are independent and identically distributed (i.i.d.) , with
\begin{equation}\label{eq:PQi}
P(Q_i=q)=\frac{\lambda(q)}{\Lambda}.
\end{equation}
The i.i.d. property of the $Q_i$ follows directly from the photons being modeled as a Poisson point process: the independence of the photons positions $Q_i$ follows from the background photons being generated independently, while their identical distribution follows from the fact that each photon is drawn from the same spatial background profile $\lambda(q)$.
The i.i.d. property propagates to the random variables $X_i=\mathcal K(Q_i)$ (because the same deterministic kernel $\mathcal K$ is applied to each of the i.i.d. random variables $Q_i$).




%

To summarize, $S$ is a sum of i.i.d.\ random variables $X_i$, the number
$N$ of terms in the sum is Poisson distributed, and $N$ is independent of the individual photon positions. By definition (\S\ref{sec:compound_def}), $S$
is therefore a \emph{compound Poisson random variable}.

\subsection{The probability distribution of $S$}

The advantage of recognizing $S$ as a compound Poisson random variable is that its probability distribution is known. \cite{Feller1968} (Chapter XII, \S 2), showed that if $n$ is Poisson distributed (with expectation $\Lambda$) and $\{X_i\}$ are $n$ i.i.d. random variables with the common probability distribution $p_X$, then their random sum (Equation~\ref{eq:compound_def}) has the probability distribution

\begin{equation}\label{eq:Feller}
\mathcal{P}_{\mathcal H_0}(s)
=
\sum_{n=0}^{\infty}
P_{\rm pois}(n)\,
p_X^{(*n)}(s),
\end{equation}


where $p_X^{(*n)}$ denotes the $n$-fold convolution of $p_X$ with
itself,
\begin{equation}
p_X^{(*n)}
=
\underbrace{p_X * p_X * \cdots * p_X}_{n\ {\rm times}},
\end{equation}
with $p_X^{(*1)}=p_X$ and $p_X^{(*0)}=\delta_0$, where
$\delta_0$ denotes a unit probability mass at zero. 

In \S\ref{sec:method}, we give a literature review of the use of compound variables or equivalent formalisms in other Astronomy-related problems and in Appendix~\ref{sec:appendix_feller}, we rederive this result in our specific context. Here, we simply use this result.  Conditioned on $N_{\rm tot}=n$, we have shown that the statistic
$S$ can be written as a sum $X_1+\cdots+X_n$ where the $\{X_i\}$ are i.i.d. with common probability distribution $p_X$. The total number $N_{\rm tot}$ of photons hitting the Kernel support is itself Poisson-distributed, and so $S$ meets the definition of the random sum, which, according to the result by \cite{Feller1968} means that $\mathcal{P}_{\mathcal{H}_0}$ is given by Equation~\ref{eq:Feller}. 
%
Therefore, what is left to evaluate is $p_X$.

\subsection{The single-photon distribution $p_X$}
\label{sec:single_photon}

Let $X_i \equiv \mathcal{K}(Q_i)$ denote the contribution of the $i$-th background photon to the statistic
$S$, where $Q_i$ is the pixel in which that photon is detected.
Therefore, the distribution of $X_i$ is determined by the spatial
distribution of the background photons (through $Q_i$) and by the kernel values.

Equation~\ref{eq:PQi} gives the distribution of the position of a background photon within the spatial kernel support (conditional on the occurrence of a background event within this region):
It follows that the single-photon distribution is
\begin{equation}
p_X(x)
\equiv
P(X_i=x)
=
\frac{1}{\Lambda}
\sum_q
\lambda(q)\,
\mathbb{1}\!\left\{\mathcal{K}(q)=x\right\},
\label{eq:one_photon_general}
\end{equation}
where $\mathbb{1}\{\cdot\}$ denotes the indicator function, equal to unity
when the condition within the braces is satisfied and zero otherwise.

For a spatially uniform background, $\lambda(q)$ is constant over the
$N_{\rm pix}$ pixels in the kernel support. Equation~\ref{eq:one_photon_general}
then reduces to
\begin{equation}
p_X(x)
=
\frac{1}{N_{\rm pix}}
\sum_{j=1}^{N_{\rm pix}}
\mathbb{1}\!\left\{\mathcal{K}_j=x\right\}
=
\frac{\#\{j:\mathcal{K}_j=x\}}{N_{\rm pix}},
\label{eq:one_photon}
\end{equation}
where $\#\{\cdot\}$ denotes the number of elements satisfying the condition
within the braces.

Thus, for a uniform background, $p_X$ is simply the histogram of the kernel
values normalized by $N_{\rm pix}$: each background photon is equally likely
to land on any pixel, and $p_X(x)$ is the fraction of pixels whose kernel
value is equal to $x$. For a spatially varying background, the same
construction becomes a weighted histogram, with each kernel value weighted
by the expected background counts in the corresponding pixel. An example
for the uniform-background case is shown in the lower-right panel of
Figure~\ref{fig:illustration}.

\subsection{Compound-variable formalism in astronomy}

Compound distributions, or equivalent formulations based directly on generating functions, have previously been used in astronomy. In this section, we briefly review these approaches and clarify the specific features of the method proposed here.

In the FFT implementation presented in \S\ref{sec:fft}, the compound Poisson distribution is expressed in terms of its characteristic function. This equivalent formalism was used by \citet{Scheuer1957} and \citet{Condon1974} in the context of radio astronomy. In their work, the individual random contributions $X_i$ are the instrumental responses produced by individual radio sources, and their sum is the total measured deflection.

More recently, a closely related formalism based on probability-generating functions was used by \citet{Malyshev2011}. Although the authors did not formulate their calculation explicitly in terms of summed random variables $X_i$, the formalism they used can be interpreted in this way: each $X_i$ is the number of photons contributed by an individual $\gamma$-ray source to a given pixel, and their sum $S$ is the total number of photons measured in that pixel. \citet{Collin2022} used the compound-Poisson formalism explicitly to construct likelihoods for unresolved point-source populations. As in \citet{Malyshev2011}, the Poisson-distributed events are astrophysical sources, and each source contributes a random number of photons to an image bin.

In contrast, in the work presented here the Poisson-distributed events are the photons themselves (and not the astrophysical sources). Each photon contributes a random weight $X_i=\mathcal{K}(Q_i)$, determined by its position $Q_i$ and by the linear-filter kernel $\mathcal{K}$. The resulting compound variable, $S=\sum_i X_i$, is the detection statistic (and more generally any linear statistic) whose distribution and extreme-tail survival probability determine the false-alarm probability and the corresponding detection threshold.

\section{Methods}\label{sec:method}
Compound variables have been extensively studied, and several strategies were proposed to compute their probability distributions. \citealt{EmbrechtsFrei2009} and \citealt{Shevchenko2010} make a comparison of these methods in the context of risk theory. Here we present two of these implementations. In the appendix \ref{appendix:enumeration}, we also present an equivalent, computationally heavier method which is not based on the compound variable formalism presented in \S\ref{sec:compound_solution_section} but rather on enumerating all the configurations of events (see the first approach in \S\ref{sec:two_approaches}).

\subsection{The FFT Implementation Using the Characteristic Function}
\label{sec:fft}

\subsubsection{Derivation}

The probability distribution $\mathcal{P}_{\mathcal{H}_0}(s)$ of the
statistic $S$ under the null hypothesis $\mathcal{H}_0$ is given by the
compound Poisson distribution derived in
Section~\ref{sec:compound_solution_section}
(Equation~\ref{eq:Feller}). Since this distribution is expressed
as a Poisson-weighted sum of convolution powers of the one-photon
probability distribution, it can be evaluated efficiently using the
convolution theorem,
\begin{equation}
\label{eq:convolution_theorem}
\mathcal{F}\!\left(p_X^{(*n)}\right)
=
\left[\mathcal{F}(p_X)\right]^n,
\end{equation}
where $\mathcal{F}$ denotes the Fourier transform, $p_X$ is the
probability distribution of the contribution $X$ of a single background
photon to the statistic, and $p_X^{(*n)}$ denotes the $n$-fold
convolution of $p_X$ with itself. Thus, repeated convolutions in the
statistic domain are replaced by pointwise multiplications in Fourier
space\footnote{For small stamps, direct implementation of convolution may be faster than using FFT.}.

This already provides an efficient way to evaluate the sum in
Equation~\ref{eq:Feller}. An even more efficient implementation
follows from the fact that the compound Poisson distribution has the
closed-form characteristic function (\citealt{Feller1968})
\begin{equation}
\label{eq:charfun_fft}
\varphi_S(\omega)
=
\exp\!\left[
\Lambda\left(
\varphi_X(\omega)-1
\right)
\right],
\end{equation}
where $\Lambda$ is the expected total number of background photons within the
support of the filter, $i$ is the imaginary unit, $\omega$ is the Fourier frequency, and
$\varphi_X(\omega)$ is the characteristic function of the one-photon
contribution $X$. The latter is defined as
\begin{equation}
\label{eq:charfun_X}
\varphi_X(\omega)
\equiv
\mathbb{E}\!\left[e^{i\omega X}\right]
=
\sum_x p_X(x)e^{i\omega x},
\end{equation}
where $\mathbb{E}$ is the expectancy value, $p_X(x)=P(X=x)$ is the probability mass function of $X$, and the
sum extends over all possible values $x$ of the one-photon contribution.

Equation~\ref{eq:charfun_fft} therefore provides the characteristic
function of the complete compound Poisson distribution directly, without explicitly summing over the number of background photons or computing successive convolution powers. In particular, this formulation allows to calculate $\mathcal{P}_{\mathcal{H}_0}(s)$ using a single Fourier transform (Equation~\ref{eq:charfun_X}), a pointwise complex exponential (Equation~\ref{eq:charfun_fft}), and a single inverse Fourier transform (to recover $\mathcal{P}_{\mathcal{H}_0}(s)$ from Equation~\ref{eq:charfun_fft}), eliminating from the implementation the dependency over the photon
number $n$ (Equation~\ref{eq:convolution_theorem}).
The numerical
implementation of these operations using the discrete Fourier transform
is described in the rest of this section.

\subsubsection{Binning the single-photon distribution}\label{sec:binning_px}
We showed in~\S\ref{sec:single_photon} that $p_X$, the single-event term in Equation~\ref{eq:Feller}, can be seen as the normalized histogram of kernel values. The first step of the implementation is to discretize it onto a lattice with spacing $\Delta s$: each kernel value $\mathcal{K}(q_\ell)$ is assigned to a bin whose index is
\begin{equation}
b_\ell
=
\mathrm{round}\!\left(
\frac{\mathcal{K}(q_\ell)}{\Delta s}
\right).
\end{equation}
Here, $\ell=1,\ldots,N_{\rm pix}$ indexes the pixels in the kernel support,
$N_{\rm pix}$ is the total number of such pixels, and $b_\ell$ is the bin
index associated with the kernel value at pixel $q_\ell$.
By default, $\Delta s=0.01$, and it is then refined if needed, as described
in~\S\ref{sec:validation}.

The binned version of Equation~\ref{eq:one_photon} gives the probability
associated with bin $j$ as
\begin{equation}
\label{eq:single_histo_binned}
p_X[j]
=
\frac{\#\{\,\ell:b_\ell=j\,\}}{N_{\rm pix}},
\end{equation}
with
\begin{equation}
    \sum_j p_X[j] = 1.
\end{equation}
where $j$ denotes the bin index and
$\#\{\,\ell:b_\ell=j\,\}$ denotes the number of kernel values assigned to
bin $j$ (see the lower left panel of Figure~\ref{fig:illustration}).

\subsubsection{Characteristic-function evaluation}

The characteristic function of the binned one-photon distribution is
\begin{equation}\label{eq:ft_px}
\phi_X(\omega)
=
\sum_j
p_X[j]\,
e^{i\omega j\Delta s},
\end{equation}
where $\Delta s$ is the bin width and $i$ is the imaginary unit.
Evaluating this expression at the discrete Fourier frequencies
\begin{equation}\label{eq:fft_freq}
\omega_k
=
-\frac{2\pi k}{N_{\omega}\Delta s},
\qquad
k=0,\ldots,N_{\omega}-1,
\end{equation}
where $k$ denotes the Fourier-frequency index and $N_{\omega}$ is the FFT
length (i.e. the number of discrete grid points used to represent the distribution and, equivalently, the number of Fourier frequencies evaluated by the FFT), gives
\begin{equation}\label{eq:DFT}
\phi_X[k]
=
\phi_X(\omega_k)
=
\sum_j
p_X[j]
\exp\!\left(
-\frac{2\pi i k j}{N_{\omega}}
\right),
\end{equation}
which is precisely the discrete Fourier transform (DFT) of the binned
probability mass function. In practice, this transform is evaluated
efficiently using the Fast Fourier Transform (FFT).

Using Equation~\ref{eq:charfun_fft}, the characteristic function of the
compound Poisson distribution is then computed as
\begin{equation}\label{eq:exponent}
\phi_S[k]
=
\exp\!\left[
\Lambda
\left(
\phi_X[k]-1
\right)
\right].
\end{equation}

Finally, the probability mass function of the matched-filter statistic is
obtained by a single inverse Fourier transform,
\begin{equation}\label{eq:inverse_fft}
\mathcal P_{\mathcal H_0}[j]
=
\mathcal F^{-1}
\!\left(
\phi_S
\right)[j].
\end{equation}

The result is real in exact arithmetic. In the numerical implementation, we
retain its real part and clip small negative values arising from floating-point
round-off.

\subsubsection{Choice of the FFT grid}\label{sec:finite_grid}

Although the compound Poisson distribution has infinite support, the
probability of observing very large photon numbers decreases rapidly.
Since the discrete Fourier transform (Equation~\ref{eq:DFT}) can only be applied to finite arrays, we truncate the computational domain so that the neglected probability is smaller than a prescribed numerical tolerance. The FFT grid is then chosen large enough that no probability within the retained range of possible values of $S$ wraps around because of the FFT's implicit periodic boundary conditions.

Specifically, we determine an integer \(n_{\rm alias}\) satisfying
\begin{equation}
P(n>n_{\rm alias})
<
\epsilon_{\rm alias},
\qquad
n\sim{\rm Poisson}(\Lambda),
\end{equation}
where $\epsilon_{\rm alias}$ is a user-defined tolerance (by default,
$\epsilon_{\rm alias}=10^{-12}$), which can then be refined if needed, as
explained in~\S\ref{sec:validation}. Therefore, the total number of photons
satisfies $n\le n_{\rm alias}$ with probability at least
$1-\epsilon_{\rm alias}$.

Let \(j_{\min}\) and \(j_{\max}\) denote the minimum and maximum occupied
indices of the one-photon probability mass function $p_X$.
Since each photon contributes an index between these two limits, any
realization containing at most \(n_{\rm alias}\) photons must lie within
the interval
\begin{equation}
j_{\rm low}
=
\min(0,n_{\rm alias}j_{\min}),
\qquad
j_{\rm high}
=
\max(0,n_{\rm alias}j_{\max}).
\end{equation}

The FFT is used to evaluate the discrete Fourier transform (DFT), which
interprets its input as one period of a periodic sequence. Consequently,
the transformed distribution is assumed to satisfy periodic boundary
conditions. If the computational grid does not fully cover the interval
\([j_{\rm low},j_{\rm high}]\), probability extending beyond one end of
the array reappears at the opposite end because of this periodic
extension (circular aliasing). We therefore choose the computational
grid to contain at least
\begin{equation}
N_{\rm grid}
=
j_{\rm high}-j_{\rm low}+1
\end{equation}
points, and set the FFT length to the smallest power of two satisfying
\begin{equation}
N_{\omega}
\ge
N_{\rm grid}.
\end{equation}

With this choice, no circular aliasing occurs within the retained
support, and the only approximation is the omission of the Poisson tail
$P(n>n_{\rm alias})$, whose total probability is bounded by
$\epsilon_{\rm alias}$.

Finally, the discrete indices are converted back into values of the
matched-filter statistic according to
\begin{equation}
s_j=j\,\Delta s,
\end{equation}
and the probability density corresponding to
\(\mathcal P_{\mathcal H_0}\) is approximated by
\begin{equation}
\mathcal P_{\mathcal H_0}(s_j)
\cong
\frac{\mathcal P_{\mathcal H_0}[j]}
{\Delta s}.
\end{equation}
That is, $\mathcal{P}_{\mathcal H_0}[j]$ is the probability associated with
the interval
\begin{equation}
[s_j-\Delta s/2,s_j+\Delta s/2].
\end{equation}

\subsubsection{Validation}\label{sec:validation}

The characteristic-function method uses: (i) a binned version of $p_X$ (\S\ref{sec:binning_px}) and (ii) a finite FFT grid (\S\ref{sec:finite_grid}).
To verify that the inferred detection threshold is insensitive to both these approximations, the calculation is repeated using
(i) a finer binning $\Delta s$, and
(ii) a stricter aliasing tolerance $\epsilon_{\rm alias}$ corresponding to a larger FFT support.
The resulting thresholds and survival functions, defined as $\overline{\mathcal P}_{\mathcal H_0}(s)=P(S\ge s \,|\,\mathcal{H}_0)$, are then compared through the five consistency checks detailed below. 

The rational of comparing both the threshold and the survival function is that it provides a stronger
convergence test than comparing the threshold alone. Indeed, two numerical
calculations may yield nearly identical threshold values while producing
noticeably different survival functions in the vicinity of the threshold,
indicating that the result is still sensitive to the numerical parameters.

First, the threshold must be stable under refinement of the binning.
If \(S_{\rm th}\) and \(S_{\rm th}^{(\Delta)}\) denote the thresholds
obtained with the nominal and refined bin widths, respectively, we require
\begin{equation}
\left|
S_{\rm th}^{(\Delta)}
-
S_{\rm th}
\right|
<
\epsilon_S,
\end{equation}
where \(\epsilon_S\) is the adopted threshold tolerance (specified by the user or set to $\Delta s=0.01$ by default).

Second, the survival function must also remain stable in the vicinity of the target false-alarm probability \(\gamma\). After interpolating the survival
function obtained on the coarse grid onto the refined grid, we require
\begin{equation}
\max_{s\in\mathcal R_\gamma}
\left|
\overline{\mathcal P}_{\mathcal H_0}^{(\Delta)}(s)
-
\overline{\mathcal P}_{\mathcal H_0}(s)
\right|
<
\max\!\left(
\epsilon_{\rm rel}\gamma,
\epsilon_{\rm abs}
\right),
\end{equation}
where \(\mathcal R_\gamma\) denotes a region surrounding the threshold, and
$\epsilon_{\rm rel}$ and $\epsilon_{\rm abs}$ are adjustable relative
and absolute tolerances (with default values $\epsilon_{\rm rel}=0.5$,
and $\epsilon_{\rm abs}=1\times{10}^{-3}$). Scaling the survival-function tolerance with the target false-alarm probability $\gamma$ ensures that the required numerical accuracy is of the same order as the rarity of the event being evaluated.

Third, the Poisson probability associated with photon numbers beyond the FFT
support must be negligible compared to the target false-alarm probability.
If $n_{\rm alias}$ denotes the photon number used to determine the FFT
support, we require
\begin{equation}
P(n>n_{\rm alias})
<
a_{\rm alias}\,\gamma,
\end{equation}
where \(a_{\rm alias}\ll1\) is a user-defined tolerance factor (with default value $0.01$). This
ensures that the probability mass potentially affected by circular aliasing is much smaller than the false-alarm probability being estimated.

Fourth, the threshold must also be stable when the aliasing tolerance is
made more stringent. If
\(S_{\rm th}^{({\rm alias})}\)
denotes the threshold obtained using the refined aliasing tolerance, we
require
\begin{equation}
\left|
S_{\rm th}^{({\rm alias})}
-
S_{\rm th}
\right|
<
\epsilon_S.
\end{equation}

Finally, the survival function must remain stable under the same refinement
of the aliasing tolerance:
\begin{equation}
\max_{s\in\mathcal R_\gamma}
\left|
\overline{\mathcal P}_{\mathcal H_0}^{({\rm alias})}(s)
-
\overline{\mathcal P}_{\mathcal H_0}(s)
\right|
<
\max\!\left(
\epsilon_{\rm rel}\gamma,
\epsilon_{\rm abs}
\right).
\end{equation}

If either of the alias-related checks fails, the aliasing tolerance
\(\epsilon_{\rm alias}\) is reduced, thereby increasing the FFT support.
Similarly, if either discretization check fails, the bin width
\(\Delta s\) is reduced. This adaptive refinement procedure is repeated
until all five consistency checks are satisfied or a prescribed maximum
number of refinement iterations is reached (by default, $10$).

\subsubsection{Computational complexity}

Constructing the one-photon histogram requires
\(\mathcal O(N_{\rm pix})\) operations.
The characteristic-function implementation then performs:

\begin{enumerate}
\item one FFT of the one-photon distribution (Equation~\ref{eq:DFT}),
\item one pointwise complex exponential (Equation~\ref{eq:exponent}),
\item one inverse FFT (Equation~\ref{eq:inverse_fft}).
\end{enumerate}

The overall computational complexity is therefore
$\mathcal O
\left(
N_{\rm pix}
+
N_{\omega}\log N_{\omega}
\right),
$
where $N_{\omega}$ is the FFT length (see Equation~\ref{eq:fft_freq}),
with a memory requirement of
$
\mathcal O(N_{\omega}).
$

\subsubsection{Catastrophic cancellations}\label{sec:cata}
A practical limitation of the FFT approach presented above is the loss of numerical accuracy in the extreme tail of the distribution due to {\it catastrophic cancellation} in the inverse Fourier transform. The compound-Poisson characteristic function can be evaluated efficiently in Fourier space, after which the probability mass function is recovered through an inverse FFT. Recovering very small probabilities requires the inverse transform to combine complex Fourier components whose contributions nearly cancel. In finite-precision arithmetic, the residual after this cancellation can become comparable to floating-point roundoff error. Consequently, the far-tail probability distribution develops small oscillations or even negative values, which are visible in the example of Figure~\ref{fig:compare_fft}. In the next section, we present an alternative to this implementation which avoids this issue.

\subsection{Panjer-Recursion Implementation}
\label{sec:panjer}


The Panjer recursion is a classical computational method for evaluating compound probability distributions. It was introduced in its
general recursive form by \cite{panjer1981recursive}. The idea is that each
probability is constructed recursively from previously computed 
probabilities and the one-photon probability mass function
$p_X$. It does not rely on the subtraction of nearly equal
oscillatory Fourier contributions and is not subject to the same
catastrophic-cancellation mechanism.

Our Panjer recursion implementation use exactly the same discretized
one-photon distribution $p_X[j]$ as in \S\ref{sec:binning_px}. The Panjer implementation requires the discretized one-photon
distribution to have non-negative support,
\begin{equation}
    b_l \geq 0.
\end{equation}
Not that when the kernel $\mathcal{K}$ is the PSF, this criterion is naturally verified, but this may not be the case with other kernels such as the Mexican hat wavelet used e.g. by {\tt wavdetect}, for which negative lattice indices may be present, in which case another implementation, such as the characteristic-function option (\S\ref{sec:fft}) must instead be used.

\subsubsection{Panjer recursion}

Let $\mathcal P_{\mathcal H_0}[j]$ denote the probability that the
background-only matched-filter statistic occupies lattice bin $j$. For a
compound Poisson distribution with rate $\Lambda$ and one-photon probability
mass function $p_X[j]$, the Panjer recursion gives
\begin{equation}
    \mathcal P_{\mathcal H_0}[0]
    =
    \exp\left[
        \Lambda\left(p_X[0]-1\right)
    \right],
\end{equation}
and, for $j\geq1$,
\begin{equation}
    \mathcal P_{\mathcal H_0}[j]
    =
    \frac{\Lambda}{j}
    \sum_{k=1}^{\min(j,j_{\max})}
    k\,p_X[k]\,
    \mathcal P_{\mathcal H_0}[j-k],
    \label{eq:panjer_recursion}
\end{equation}
where $j_{\max}$ is the largest occupied index of the one-photon probability
mass function.

The factors $k\,p_X[k]$ are precomputed,
and $\mathcal P_{\mathcal H_0}[j]$ is obtained successively from previously
calculated values $\mathcal P_{\mathcal H_0}[j-k]$. After completing the
recursion, the resulting probability mass function is normalized so that
\begin{equation}
    \sum_j \mathcal P_{\mathcal H_0}[j] = 1.
\end{equation}

As in the FFT implementation, the discrete indices are converted back into
values of the matched-filter statistic according to
\begin{equation}
    s_j = j\,\Delta s,
\end{equation}
and the corresponding probability density is approximated by
\begin{equation}
    \mathcal P_{\mathcal H_0}(s_j)
    \approx
    \frac{\mathcal P_{\mathcal H_0}[j]}{\Delta s}.
\end{equation}

\subsubsection{Stopping criterion}

The numerical calculation of  $\mathcal{P}_{\mathcal H_0}$ only needs to include photon counts up to a sufficiently large
value $n_{\max}$. The implementation chooses $n_{\max}$ such that the probability omitted by truncating the calculation is lower than a tolerance limit $\epsilon_{\rm count}$ set by the user:
\begin{equation}
    P(n > n_{\max}) \leq \epsilon_{\rm count},
    \qquad
    n\sim\operatorname{Poisson}(\Lambda),
\end{equation}

  and $n_{\max}$ is obtained from the inverse Poisson survival function.

If $j_{\max}$ is the largest occupied index of the one-photon probability
mass function $p_X$, the largest index of the compound distribution that must
be represented is
\begin{equation}
    j_{\rm high} = n_{\max}j_{\max}.
\end{equation}
Any realization producing a statistic beyond $j_{\rm high}$ must therefore
contain more than $n_{\max}$ photons, so that
\begin{equation}
    P(S > j_{\rm high}\Delta s)
    \leq P(n>n_{\max})
    \leq \epsilon_{\rm count}.
\end{equation}
Thus, the extent of the computed distribution is not decided in advance, but rather controlled directly through a bound on the omitted probability.

The target tail probability $\gamma$ used to determine the detection
threshold is distinct from the truncation tolerance $\epsilon_{\rm count}$.
After the recursion has been evaluated, the survival function is computed as
\begin{equation}
    {\rm SF}(S_j)
    =
    P(S\geq S_j\mid\mathcal H_0)
    =
    \sum_{\ell=j}^{j_{\rm high}}
    \mathcal P_{\mathcal H_0}[\ell],
\end{equation}
and the detection threshold is the smallest $S_j$ for which
\begin{equation}
    {\rm SF}(S_j)\leq\gamma.
\end{equation}
To ensure that truncation of the Poisson-count distribution does not limit
the accuracy of the tail probability of interest, $\epsilon_{\rm count}$
is chosen sufficiently smaller than $\gamma$.

\subsubsection{Validation}

The exact same five validation steps as in the FFT implementation (\S\ref{sec:validation}) are performed, the analogous to $\epsilon_{alias}$ being the photon count $\epsilon_{\rm count}$ determining how far the Panjer recursion must be evaluated. The Panjer threshold is declared validated only if all five conditions are
satisfied simultaneously. If either of the count-tail checks fails,
$\epsilon_{\rm count}$ is reduced, thereby increasing the maximum photon
number included in the recursion. Similarly, if either discretization check
fails, the bin width $\Delta s$ is reduced. This adaptive refinement procedure
is repeated until all five consistency checks are satisfied or the prescribed
maximum number of refinement iterations is reached.

\subsubsection{Computational complexity}

For the Panjer implementation, constructing the one-photon histogram
similarly to the FFT case requires $\mathcal{O}(N_{\rm pix})$ operations. The recursion
then evaluates $\mathcal{P}_{\mathcal{H}_0}$ sequentially over
$N_{\rm grid}$ bins. For each output bin, the recursion involves a sum
over the $N_X$ occupied bins of the one-photon distribution $p_X$, giving
an overall computational complexity of
\begin{equation}
\mathcal{O}\!\left(
N_{\rm pix}
+
N_{\rm grid}N_X
\right),
\end{equation}
where $N_X$ is the number of occupied bins of $p_X$. Since $p_X$ is
constructed from $N_{\rm pix}$ kernel values, $N_X\leq N_{\rm pix}$, and
the computational complexity is therefore bounded by
\begin{equation}
\mathcal{O}\!\left(
N_{\rm pix}
+
N_{\rm grid}N_{\rm pix}
\right).
\end{equation}
The memory requirement is $\mathcal{O}(N_{\rm grid})$.

\section{Code}
\label{sec:code}

A {\tt Python} implementation of this tool is released to the community alongside the publication of this paper through MTS's \href{https://github.com/maayane/}{Github page}. Information on how to install and run the pipeline, as well as on its architecture, will be provided on this page.

A MATLAB implementation is available as part of the AstroPack/MAATv2 package\footnote{\url{https://github.com/EranOfek/AstroPack}} \citep{Ofek2014_MAAT, Ofek+2023PASP_LAST_PipeplineI}.
The functions: {\tt poissThresholdSimulations}, {\tt poissThresholdFFT}, {\tt poissThresholdPanjer}, and {\tt unitTest_poissThreshold} are available in the {\tt imUtil.poissNoise} package.
\section{Tests}\label{sec:tests}

\begin{figure*}
    \centering
    \includegraphics[width=0.48\linewidth]{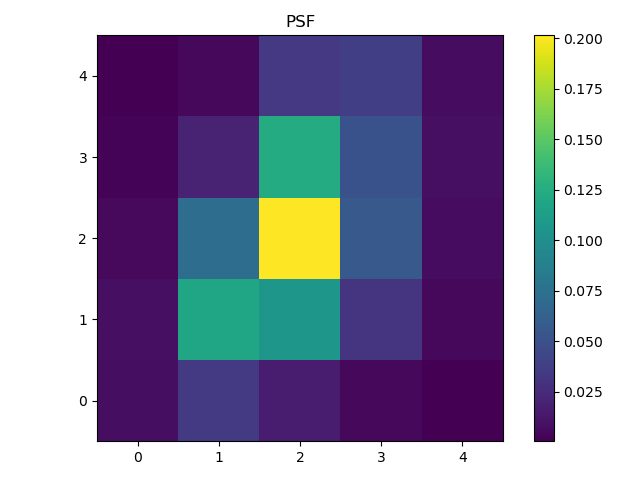}
    \includegraphics[width=0.48\linewidth]{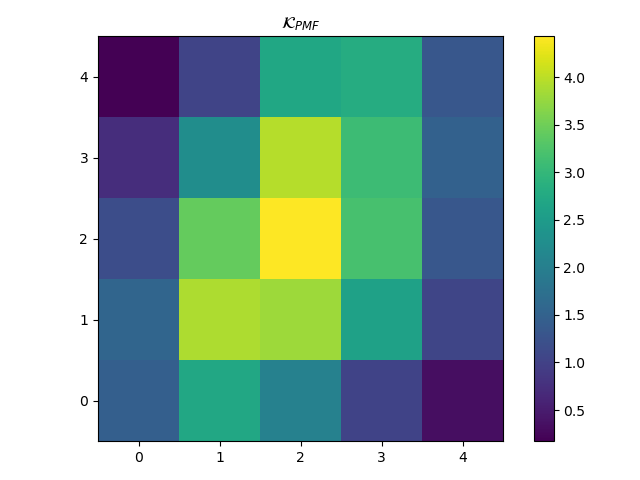}
     \includegraphics[width=0.45\linewidth]{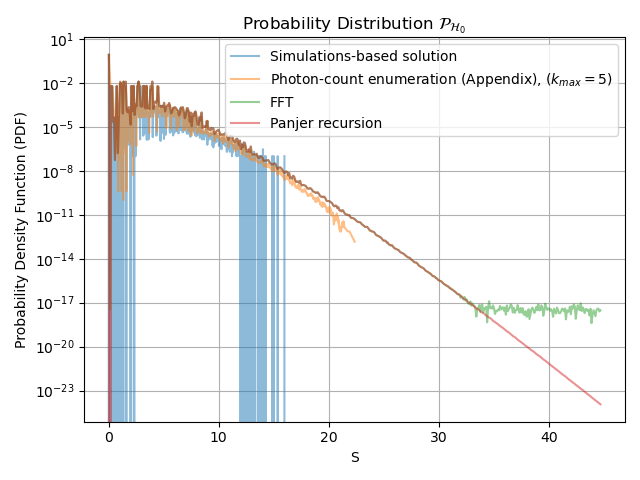}
     \includegraphics[width=0.45\linewidth]{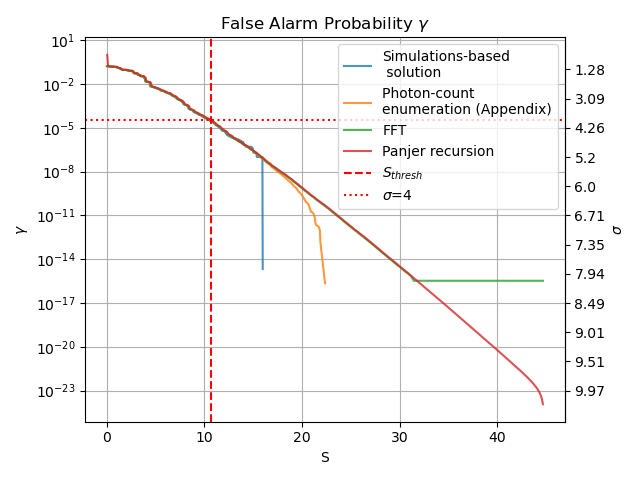}    
    \caption{{\bf Implementing the method in a 2D image.} Top right: a PSF stamp. Top left: The optimal PMF kernel corresponding to this stamp, a background value $B=7.3\times 10^{-3}$ and a flux $F=3$ (see Equation~\ref{eq:kpmf}).
    Lower Left: the probability distribution of $S$ calculated for this kernel: with $10^7$ simulations (blue), with the enumeration of photons configurations described in the appendix (orange), with the FFT implementation described in \S\ref{sec:fft} (green) and with the Panjer recursion implementation presented in \S\ref{sec:panjer} (red). The oscillations due to catastrophic cancellations in the FFT implementation (\S\ref{sec:cata}) are well visible.  Lower Right: the false-alarm probability $\gamma$, from which the threshold $S_{\rm th}$ above which a detection is declared is calculated. The four methods of calculation lead to a value of $S_{\rm th}$ that differs by only $1\%$, but the FFT and Panjer methods takes significantly less time to run ($30\,\rm msec$ versus $1\,\rm sec$ for the enumeration solution and $12\,\rm sec$ for the simulations-based solution).}
    \label{fig:compare_fft}
\end{figure*}

In Figure~\ref{fig:compare_fft}, we show the probability distribution and the false alarm probability as a function of the statistic $S$, computed using each of the four following methods: (1) Sampling $\mathcal{P}_{\mathcal H_0}$ by producing $10^8$ simulations of pure background images (2) the solution, presented in \S\ref{appendix:enumeration}, based on enumerating the occupancy vectors $\boldsymbol{m}$, with a total number of photons $n_{\rm tot}=5$ (3) the FFT-based implementation presented in \S\ref{sec:fft} and (4) the Panjer recursion implementation presented in \S\ref{sec:panjer}. 

The last solution, in addition to being much faster, allows to push the calculation to much higher values of the total number of photons in the background image, whereas the calculation based on the enumeration of the configuration becomes prohibitively time-consuming beyond $n_{\rm tot}=5$.

For the \(5\times5\) kernel stamp used in Figure~\ref{fig:compare_fft}, the FFT-based implementation computes \(\mathcal{P}_{\mathcal{H}_0}\) and the corresponding threshold \(S_{\rm th}\) in only \(30\) ms on a single core of a 2.3 GHz 8-core Intel Core i9 processor, compared with \(0.5\) s for the enumeration method and \(10\) s for the simulations.
We note that the FFT and Panjer implementations still provides good results even for stamps as larger than $100\times100$, a size for which the computational cost of the two other methods is prohibitive. 

\section{Conclusions}\label{sec:conclusion}

We present a new method for computing the false-alarm probabilities and the detection thresholds for arbitrary linear filters in the low-count regime, where the background noise follows Poisson statistics. The method is accurate, independent of the choice of filter (it works for any linear detection statistic) and avoids the need for computationally expensive simulations.

By recognizing the detection statistic as a {\it compound Poisson random variable}, the full probability distribution of the detection statistic can be shown to be completely determined by two quantities: the total number of background photons and the one-photon probability distribution $p_X$. As a result, it
can be computed efficiently using the FFT, or the Panjer transform, yielding orders-of-magnitude improvements in computational speed while maintaining high accuracy. This solution enables reliable evaluation of extremely low false-alarm probabilities that are often impractical to estimate by simulation alone. Although motivated by astronomical source detection, this method is broadly applicable to any detection problem involving linear statistics and Poisson-distributed events. Python and MATLAB implementations are provided to facilitate its adoption in a wide range of scientific applications.



\acknowledgments
This paper is dedicated to the memory of Lucien Lazare.

We thank an anonymous referee of \cite{Soumagnac-PMF}, whose comments led us to realize the importance of deriving the analytical solution presented in this paper.

ChatGPT (OpenAI) was used during the development of this work to help bridge gaps in our familiarity with methods used in Risk theory for computing compound probability distributions and with the associated literature, e.g. \cite{Feller1968}.

This research was supported by the Israeli Science Foundation (grant Nos. 2068/22 and 2751/22)

This research has made use of data obtained from the Chandra Data Archive and the Chandra Source Catalog, both provided by the Chandra X-ray Center (CXC).

This research used resources of the National Energy Research Scientific Computing Center (NERSC), a Department of Energy Office of Science User Facility using NERSC award DDR-ERCAP-0029239.

E.O.O. is grateful for support by grants from the Willner Family Leadership Institute, Madame Olga Klein-Astrachan, André Deloro Institute, Schwartz/Reisman Collaborative Science Program, Paul and Tina Gardner, The Norman E Alexander Family Foundation ULTRASAT Data Center Fund, Jonathan Beare, Israel Science Foundation, Minerva, BSF, BSF-transformative, and the Weizmann-UK.

\appendix

\section{Calculating the false alarm rate via Photon-Count Enumeration}
\label{appendix:enumeration}





In the paper, the probability distribution of the detection statistic $S$ (Equation~\ref{eq:Feller}) was obtained through the {\it second} of the two representations described in \S\ref{sec:two_approaches}, i.e. by constructing $\mathcal{P}_{\mathcal{H}_0}(s)$ through the contributions of individual {\it photons}. Within this representation, $S$ meets the definition of a compound Poisson random variable, of which the probability distribution is well known \citep{Feller1968}. Here, we present another way to reach the same result, through the {\it first} of the two representations described in \S\ref{sec:two_approaches}. The idea is to enumerate all the possible realizations of the occupancy vector $\boldsymbol{M_{\rm b}}$ and to
construct the corresponding distribution of $S$ by assigning a probability to each possible occupancy configuration. In Appendix~\ref{appendix:equivalence}, we demonstrate formally the equivalence between these two representations.

\subsubsection{Derivation}
We start with equation~\ref{eq:filter_simple}:
\begin{equation}\label{eq:statistic}
S=\sum_{q=1}^{N_{\rm pix}} M_{\rm b}(q)\mathcal{K}(q).
\end{equation}

Let $\boldsymbol{M_{\rm b}}=(M_{\rm b,1},M_{\rm b,2},\ldots,M_{\rm b,N_{\rm bpix}})$
denote the random occupancy vector, where $M_{\rm b,j}\equiv M_{\rm b}(q_j)$ is the random number of
counts in pixel $j$ in the background-only image, and $N_{\rm pix}$ is the number of pixels\footnote{Here
the term pixel may refer to any dimension.}. A particular realization of
$\boldsymbol{M_{\rm b}}$ is denoted by
\begin{equation}
\boldsymbol{m}=(m_1,m_2,\ldots,m_{N_{\rm pix}}).
\end{equation}

The total number of photons is the random variable
\begin{equation}
N_{\rm tot}=\sum_{j=1}^{N_{\rm pix}} M_j,
\end{equation}
whose particular realization for the occupancy vector $\boldsymbol{m}$ is
\begin{equation}\label{eq:ntot}
n_{\rm tot}(\boldsymbol{m})
=
\sum_{j=1}^{N_{\rm pix}} m_j.
\end{equation}
Thus, a realization of the occupancy vector containing $n$ photons is any
vector $\boldsymbol{m}$ satisfying
\[
\sum_{j=1}^{N_{\rm pix}} m_j=n.
\]

For a particular realization $\boldsymbol{m}$ of the occupancy vector, the
statistic from Equation~\ref{eq:statistic} is
\begin{equation}\label{eq:contributions}
S(\boldsymbol{m})
=
\sum_{j=1}^{N_{\rm pix}}
m_j\,\mathcal{K}(q_j).
\end{equation}
For example, in Figure~\ref{fig:illustration}, we show two realizations of the occupancy vector satisfying $n_{\rm tot}(\boldsymbol{m})=2$ and $N_{\rm pix}=5$:
\begin{equation}
\boldsymbol{m}
=
(0,1,0,1,0),
\end{equation}
where the two non-zero entries correspond to pixels $q_2$ and $q_4$, with $q_2\ne q_4$. For this realization,
\begin{equation}
S(\boldsymbol{m})
=
\mathcal{K}(q_2)+\mathcal{K}(q_4).
\end{equation}
Another realization with $n_{\rm tot}(\boldsymbol{m})=2$ is
\begin{equation}
\boldsymbol{m}^{'}
=
(0,0,2,0,0),
\end{equation}
where both photons fall in the same pixel $q_3$. For this realization,
\begin{equation}
S(\boldsymbol{m'})
=
2\mathcal{K}(q_3).
\end{equation}

The probability that the occupancy vector takes the realization
$\boldsymbol{m}$ is simply the product of the Poisson probabilities in each pixel,
\begin{equation}\label{eq:product}
P(\boldsymbol{M_{\rm b}}=\boldsymbol{m})
=
\prod_{j=1}^{N_{\rm pix}}
P(M_{\rm b,j}=m_j),
\end{equation}
where
\begin{equation}\label{eq:poisson_indiv}
P(M_{\rm b,j}=m_j)
=
\frac{B^{m_j}e^{-B}}{m_j!}.
\end{equation}

The probability distribution
$\mathcal{P}_{\mathcal{H}_0}(s)\equiv P(S=s\mid \mathcal{H}_0)$
is obtained by summing the probabilities of all possible realizations that
give the value $s$,
\begin{equation}\label{eq:heavy_enumeration}
\mathcal{P}_{\mathcal{H}_0}(s)
=
\sum_{\boldsymbol{m}}
P(\boldsymbol{M_{\rm b}}=\boldsymbol{m})
\,
\delta\!\left(
s-S(\boldsymbol{m})
\right),
\end{equation}
where $\delta$ denotes the delta function, and the sum runs over all
possible realizations
$\boldsymbol{m}=(m_1,\ldots,m_{N_{\rm pix}})$ of the occupancy vector,
with $m_j\ge0$. 

\subsection{Practical implementation}

Our implementation takes advantage of the fact that, when the background is
spatially uniform, all configurations sharing the same occupancy pattern
(i.e., the same partition of the total number of photons among the occupied
pixels) have the same probability. This probability therefore needs to be
calculated only once for each occupancy pattern.

For example, for $n_{\rm tot}=5$, the possible occupancy patterns are:
$5;\,4+1;\, 3+2;\,3+1+1;\, 2+2+1;\,
2+1+1+1;\, 1+1+1+1+1$.
Each pattern corresponds to many occupancy vectors which share the same probability. For example, the pattern $4+1$ includes every
configuration in which one pixel contains four photons and a second,
distinct pixel contains one photon, irrespective of the identities of these
two pixels.

For each occupancy pattern, we first compute the probability of one
representative occupancy vector as the product of the Poisson probabilities
in the individual pixels. This is the probability of all the configurations sharing this occupancy pattern. For example, for the pattern $4+1$, the probability of all the occupancy vectors corresponding to this occupancy pattern is
\begin{equation}
P_{\rm rep}(4+1)
=
P(M=4)\,
P(M=1)\,
P(M=0)^{N_{\rm pix}-2},
\end{equation}
where $P(M=k)$ is the Poisson probability of observing $k$ photons in a
given pixel.

We then calculate all values of the statistic $S$ associated with the
given occupancy pattern. For the $4+1$ pattern, these values are
\begin{equation}
S
=
4\mathcal{K}_i+\mathcal{K}_j,
\qquad i\neq j,
\end{equation}
for all pairs of distinct pixels $i$ and $j$. Each such combination is assigned the representative probability $P_{\rm rep}(4+1)$. If several combinations yield the same value of \(S\), their probabilities are summed.

In spite of the fact that probabilities must be calculted only once per occupancy pattern (and not once per occupancy vector), this solution pose a computational challenge: even for moderate values of $n_{tot}$, the number of occupancy vectors satisfying
$\sum_jm_j=n$ grows rapidly with both $n$ and $N_{\rm pix}$ and therefore the number of values of $S(\boldsymbol{m})$ which need to be calculated can quickly become prohibitive. 
%

\subsection{Computational complexity}

For a fixed total number of photons $n$, the calculation is organized by
occupancy pattern. For each pattern, the algorithm generates the allowed
combinations or permutations of occupied pixels and directly calculates the
corresponding values of the statistic $S$.

Consider an occupancy pattern with $r$ occupied pixels and multiplicities
$n_1,\ldots,n_r$, where
\begin{equation}
\sum_{\ell=1}^{r} n_\ell=n.
\end{equation}
For each allowed choice of $r$ distinct pixels, the algorithm calculates a
value of the form
\begin{equation}
S
=
\sum_{\ell=1}^{r} n_\ell \mathcal{K}_{i_\ell}.
\end{equation}
The number of such calculations scales as
$\mathcal{O}(N_{\rm pix}^{\,r})$. Since evaluating each value of $S$
requires a number of arithmetic operations that depends on $r$ but not on
$N_{\rm pix}$, the computational cost for a given occupancy pattern also
scales as $\mathcal{O}(N_{\rm pix}^{\,r})$.

The largest possible number of occupied pixels is $r=n$, corresponding to
the occupancy pattern $1+\cdots+1$, in which all $n$ photons fall in
distinct pixels. In this case, the algorithm calculates
$\binom{N_{\rm pix}}{n}$
values of $S$. For fixed $n$ and large $N_{\rm pix}$,
\begin{equation}
\binom{N_{\rm pix}}{n}
\sim
\frac{N_{\rm pix}^{\,n}}{n!},
\end{equation}
and the computational cost therefore scales as $\mathcal{O}(N_{\rm pix}^{\,n})$.

If photon numbers up to $n_{\max}$ are included, the highest-order
dependence on $N_{\rm pix}$ arises from the $n=n_{\max}$ term. For fixed
$n_{\max}$, the overall computational complexity therefore scales as $\mathcal{O}(N_{\rm pix}^{\,n_{\max}})$.

\section{Equivalence between the two representations}\label{appendix:equivalence} 
Here, we demonstrate formally the equivalence between the two representations listed in \S\ref{sec:two_approaches}, i.e. constructing $\mathcal{P}_{\mathcal{H}_0}(s)$ by enumerating occupancy 
vectors of by seeing $S$ as a compound Poisson random variable.

Decomposing the sum in Equation~\ref{eq:heavy_enumeration} according
to the total number of photons
$N_{\rm tot}=\sum_{j=1}^{N_{\rm pix}} M_j$
(i.e., applying the law of total probability), gives
\begin{equation}\label{eq:PH0}
\mathcal{P}_{\mathcal{H}_0}(s)
\;=\;
\sum_{n=0}^{\infty}
P(N_{\rm tot}=n)\;
P(S=s \mid N_{\rm tot}=n)\,.
\end{equation}
The first factor is straightforward to calculate: a sum of independent Poisson
variables is Poisson distributed with the sum of the means, so
\begin{equation}\label{eq:poiss}
P_{\rm pois}(n)
\;\equiv\;
P(N_{\rm tot}=n)
\;=\;
\frac{\Lambda^{n} e^{-\Lambda}}{n!}\,,
\qquad
\Lambda \;=\; N_{\rm pix}\,B\,.
\end{equation}
The remainder of this section is devoted to the second factor,
$P(S=s \mid n_{\rm tot}=n)$.


Since any realization $\boldsymbol{m}$ satisfying
$\sum_{j=1}^{N_{\rm pix}} m_j=n$
necessarily has
$n_{\rm tot}(\boldsymbol{m})=n$, the event
$\{\boldsymbol{M_{\rm b}}=\boldsymbol{m}\}$ is contained in the event
$\{N_{\rm tot}=n\}$. Therefore,
\begin{equation}
P\!\left(
\{\boldsymbol{M_{\rm b}}=\boldsymbol{m}\}
\cap
\{N_{\rm tot}=n\}
\right)
=
P(\boldsymbol{M_{\rm b}}=\boldsymbol{m}),
\end{equation}
and the definition of conditional probability gives
\begin{align}
P(\boldsymbol{M_{\rm b}}=\boldsymbol{m}\mid N_{\rm tot}=n)
=
\frac{
P\!\left(
\{\boldsymbol{M_{\rm b}}=\boldsymbol{m}\}
\cap
\{N_{\rm tot}=n\}
\right)}
{P(N_{\rm tot}=n)}
\nonumber \\[1ex]
=
\frac{P(\boldsymbol{M_{\rm b}}=\boldsymbol{m})}
     {P(N_{\rm tot}=n)}.
\end{align}

Substituting $P(\boldsymbol{M}=\boldsymbol{m})$ from
Equation~\ref{eq:product} and~\ref{eq:poisson_indiv} together with
Equation~\ref{eq:poiss}, yields

\begin{align}
P(\boldsymbol{m} \mid n_{\rm tot}=n)
&=
\frac{\displaystyle\prod_{j=1}^{N_{\rm pix}}
\frac{B^{m_j}e^{-B}}{m_j!}}
{\displaystyle\frac{\Lambda^{n}e^{-\Lambda}}{n!}}
\nonumber\\[1ex]
&=
\frac{B^{\sum_j m_j}e^{-BN_{\rm pix}}}
{\Lambda^n e^{-\Lambda}}
\,
\frac{n!}{\prod_{j=1}^{N_{\rm pix}}m_j!}
\nonumber\\[1ex]
&=
\frac{B^n}{(N_{\rm pix}B)^n}
\,
\frac{n!}{\prod_{j=1}^{N_{\rm pix}}m_j!}
\nonumber\\[1ex]
&=
\frac{n!}{\prod_{j=1}^{N_{\rm pix}} m_j!}
\left(\frac{1}{N_{\rm pix}}\right)^{n}.
\label{eq:multinomial}
\end{align}

This is the multinomial distribution with
$n$ trials and equal cell probabilities $1/N_{\rm pix}$ -- the
distribution obtained by throwing $n$ balls independently into
$N_{\rm pix}$ equally likely boxes, or $n$ independent pixels into $N_{\rm pix}$ equally likely pixels. In other words, the conditional distribution $P(\boldsymbol{m} \mid n_{\rm tot}=n)$ of the occupancy vector is identical to the distribution generated by assigning each of the $n$
photons independently and uniformly to one of the $N_{\rm pix}$ pixels. This brings us back to the second representation described in \S\ref{sec:two_approaches}, where $S$ is viewed as a sum of i.i.d. contributions from the $n$ photons, which allows to use the result by \cite{Feller1968} for $\mathcal{P}_{\mathcal{H}_0}$.

\section{Derivation of the Feller (1968) result for $\mathcal{P}_{\mathcal{H}_0}$}\label{sec:appendix_feller}

Here, we rederive the result by \cite{Feller1968} in our specific, photon-starved image context. We begin with Equation~~\ref{eq:PH0}:
\begin{equation}\label{eq:appendix_PH0}
\mathcal{P}_{\mathcal{H}_0}(s)
\;=\;
\sum_{n=0}^{\infty}
P(N_{\rm tot}=n)\;
P(S=s \mid N_{\rm tot}=n)\,.
\end{equation}

Conditioned on $N_{\rm tot}=n$, we have shown in \S\ref{sec:compound_proof} that the statistic
$S$ can be written as a sum $X_1+\cdots+X_n$ where the $\{X_i\}$ are i.i.d. with common probability distribution $p_X$. The probability distribution of a
sum of independent discrete random variables is the convolution of their individual probability distributions. Explicitly, for $n=2$,
\begin{align}\label{eq:nfold}
P(S=s \mid n_{\rm tot}=2)
&=
\sum_{x}
P(X_1 = x)\,P(X_2 = s-x) \notag \\
&=
\sum_{x} p_X(x)\,p_X(s-x)
\;\equiv\;
\left(p_X * p_X\right)(s)\,,
\end{align}
where the sum runs over the (finitely many) values $x$ in the support
of $p_X$: fixing the contribution $x$ of the first photon, the second
photon must contribute exactly $s-x$. Iterating this argument gives, for any $n\ge1$,
\begin{equation}\label{eq:convo_appendix}
P(S=s\mid N_{\rm tot}=n)
=
p_X^{(*n)}(s),
\end{equation}
where 
\[
p_X^{(*1)}=p_X,
\]
and the convolution powers are defined recursively by
\[
p_X^{(*(n+1))}
=
p_X^{(*n)}*p_X.
\]
The remaining case $n=0$ corresponds to no photons falling within the
kernel support. In this case,
\[
P(S=s\mid N_{\rm tot}=0)
=
\mathbb1\{s=0\},
\]
which is identified with the zeroth convolution power,
\[
p_X^{(*0)}(s)=\mathbb1\{s=0\}.
\]

In words, when no background photon falls within the kernel support, the statistic is exactly zero (this
term produces the prominent spike at $S=0$ visible in
Figure~\ref{fig:compare_fft} at low background levels.)

Substituting Equation~\ref{eq:convo_appendix} in Equation~\ref{eq:appendix_PH0} gives 
\begin{equation}\label{eq:compound_poisson_appendix}
\mathcal{P}_{\mathcal{H}_0}(s)
\;=\;
\sum_{n=0}^{\infty}
P_{\rm pois}(n)\;
p_X^{(*n)}(s)\,.
\end{equation}
This is precisely the result by \cite{Feller1968} given in Equation~\ref{eq:Feller}.

\bibliographystyle{apj} 
\bibliography{bibliograph.bib,papers}


%

\end{document}